\documentstyle[twoside,fleqn,espcrc2]{article}
\newcommand{\AmS}{{\protect\the\textfont2
  A\kern-.1667em\lower.5ex\hbox{M}\kern-.125emS}}

\title{Quantum Interference and the Trapped Bose Condensed System}

\author{Juhao Wu and A. Widom  
\address{Physics Department, Northeastern University, Boston MA 02115}
}

\begin{document}

\begin{abstract}
In experiments involving Bose condensed atoms trapped in magnetic 
bottles, plugging the hole in the bottle potential with a LASER beam 
produces a new potential with two minima, and thus a condensate order 
parameter (i.e. wave function) with two maxima. When the trapping potential 
is removed and the condensate explodes away from the trap, the two wave 
function maxima act as two coherent sources which exhibit amplitude 
interference. A simplified theoretical treatment of this experimental 
effect is provided by considering momentum distributions.
\end{abstract}

\maketitle

The two fluid\cite{1} model has long served as the standard picture  
of superfluidity in liquid $He^4$. In spite of the great success 
of the model, Landau warned\cite{2} that the view was merely quasi-classical. 
It would be desirable to have a quantum  
fluid mechanical model in which superposition of amplitudes on  
a macroscopic scale played a more central role. For the dilute Bose 
condensed gas\cite{3}, such coherence would mean that the Bose 
condensate order parameter\cite{4} $\Psi ({\bf r})$ would be required to 
maintain many of the properties normally attributed to single particle wave 
functions exhibiting amplitude superposition\cite{5}.

Recent experimental probes\cite{6}, developed for observing Bose 
condensates\cite{7} in atomic traps\cite{8},  
have provided that which is needed to go beyond the Landau 
``quasi-classical'' two fluid picture. Direct observation of amplitude 
superposition in the Bose condensate order parameter 
$\Psi ({\bf r})$ has been observed\cite{9} in mesoscopic samples.  

Consider some central features of recent experimental 
methods. To see what is involved, let us first contemplate a 
``thought experiment''. Suppose that at times $t<0^-$, 
there is but one atom localized near the origin in the region of  
a trap potential $U({\bf r})$. At time $t=0$, the trap potential is 
suddenly removed, and thereafter $t>0^+$ the atomic 
motion is described by the free particle Schr\"odinger equation 
$  
i\hbar \big( \partial \psi /\partial t)=
-(\hbar^2/2M)\nabla^2 \psi  
$.

From the viewpoint of quantum mechanics, the sudden\cite{10} removal of 
the trap potential yields an out going wave function described using 
the non-relativistic propagator\cite{11} $G({\bf r}-{\bf s},t)$. The outgoing 
wave function is determined by    
\begin{equation} 
\psi ({\bf r},t)=
\int G({\bf r}-{\bf s},t)\psi ({\bf s},0)d^3s ,
\end{equation}
where 
\begin{equation}
G({\bf r}-{\bf s},t)=\Big({M\over 2\pi i\hbar t}\Big)^{3/2}
\exp\Big({iM|{\bf r}-{\bf s}|^2\over 2\hbar t}\Big).
\end{equation}

From the viewpoint of classical mechanics, the particle exits 
the trap region at a uniform velocity along the path 
$
{\bf r}={\bf v}t
$.
Both the classical (measurement) and quantum (virtual) viewpoints 
are resolved by considering what happens after the previously trapped 
particle freely moves far away from the original trapping region. 
Employing  
$
|{\bf r}|>>|{\bf s}|
$
in Eqs.(1) and (2), one finds that 
\begin{equation}
\psi ({\bf r},t)\approx \big(M/(2\pi i\hbar t)\big)^{3/2}
e^{iS({\bf r},t)/\hbar }A(M{\bf r}/t),
\end{equation}
where $S({\bf r},t)=(M|{\bf r}|^2/2t)$ and 
\begin{equation}
A({\bf p})=\int e^{-i({\bf p\cdot s})/\hbar }\psi ({\bf s},0)d^3s.
\end{equation}
is the momentum amplitude of the particle when it was trapped by the 
potential. Eq.(3) implies that 
\begin{equation}
|\psi ({\bf r},t)|^2\approx \int |A({\bf p})|^2 
\delta \big({\bf r}-({\bf p}t/M)\big){d^3p\over (2\pi \hbar )^3}\ .  
\end{equation}

Thus, when a particle has escaped away from the trapping region (long 
after the trap potential has been removed), the probability density 
in space 
$
|\psi ({\bf r},t)|^2
$ 
can be computed {\em by averaging the classical equation of motion}  
$
{\bf r}=({\bf p}t/M)
$
over the momentum probability density    
$
|A({\bf p})|^2
$, present when the particle was in a trapped state. 

To understand recent experiments on quantum interference for trapped 
Bose condensed mesoscopic systems, one must replace the above 
single particle amplitudes with their second quantized operator 
counterparts 
\begin{equation}
\hat{\psi }({\bf r})=\int e^{i({\bf p\cdot r})/\hbar }
\hat{A}({\bf p}){d^3p\over (2\pi \hbar )^3}\ . 
\end{equation}
The many Bosons are held in the trap for times $t<0$ with an 
equilibrium momentum distribution 
\begin{equation}
N({\bf p})=\big<\hat{A}^\dagger ({\bf p})\hat{A}({\bf p})\big>.
\end{equation}
The atomic interaction potentials have importance in assuring that 
the momentum distribution $N({\bf p})$ (for the Bosons within the 
trap potential) actually correspond to thermal equilibrium. 
The experimentalist at time $t=0$ {\em suddenly turns off the 
trap potential}\cite{12} and optically monitors the particle density 
for times $t>0$. The density is determined by 
\begin{equation}
\rho ({\bf r},t)=\big<\hat{\psi }^\dagger ({\bf r},t)
\hat{\psi }({\bf r},t) \big>,
\end{equation}
{\em after} the Bosons move away from the original trapping potential 
spatial region. 

To the extent that the interaction potentials have 
a negligible effect on the particles after they explode away from the 
trapping region, the final mean density in space $\rho({\bf r},t)$ 
is related to the initial mean momentum distribution $N({\bf p})$  
via the many body version of Eq.(5); i.e. 
\begin{equation}
\rho ({\bf r},t)\approx \int N({\bf p})
\delta \big({\bf r}-({\bf p}t/M)\big){d^3p\over (2\pi \hbar )^3}\ .
\end{equation} 
The {\em ballistic} Eq.(9) has been the central theoretical tool 
employed in the recent measurements of trapped Bose condensed systems. 
{\em The momentum distribution 
of particles within the trap is probed by the spatial distribution of 
particles detected away from the trap}. 

Let us now review the notion of superposition of amplitudes as it becomes 
apparent in the classic ``two slit quantum interference experiment''
\cite{13}. 
Quantum particle diffraction experiments present us with a closely 
analogous situation regarding initial probability distributions in 
momentum and final probability distributions in space. 
In standard slit diffraction experiments, {\em the 
momentum distribution of particles just behind the slits is 
probed by the spatial distribution of the particles at the detectors 
very far behind from slits}. Once this close analogy between diffraction 
experiments and recent Bose condensed experiments is appreciated, 
it then becomes evident why these measurements have  
exhibited interference effects on the mesoscopic scale of the 
trapped Bose condensates. 

Our brief review of diffraction theory from slits is as follows: 
(i) A particle with 
high velocity $v$ is normally incident on a screen with one 
or more slits. The slits have a length scale $\bar{d}$ and 
the high incoming momentum is defined by the inequality 
$Mv>>(\hbar/\bar{d})$. 
(ii) After a time period 
\begin{equation} 
t_{s}\approx (D/v)\ \ {\rm (slit\ to\ detector\ time )},
\end{equation}
the particle hits some detector at a distance $D$ behind the slits. 
(iii) Before the particle hits the slits from incidence side of the 
screen, the momentum parallel to the screen is zero. If the particle 
is squeezed through the slits to the emission side of the screen 
at time zero, then the quantum mechanical uncertainty principle 
forces the momentum amplitude  
\begin{equation}
A(p)=\int_{slits}e^{-i(px^\prime )/\hbar}\psi (x^\prime ,0)dx^\prime .
\end{equation}
The momentum component $p$ is in the $x$-direction, and lives 
in the plane of the screen with the slits. In that plane, $p$ is 
normal to the long slit direction. 
(iv) Far behind the slits, the spatial 
distribution in $x$ near the detectors is given by a one-dimensional 
version of Eq.(5); It is   
\begin{equation}
|\psi(x,t_s)|^2 \approx \int |A(p)|^2 
\delta \big(x-(pt_s/M)\big){dp\over (2\pi \hbar )}.
\end{equation} 
From Eqs.(10) and (12) follows the general slit diffraction formula
\begin{equation}
|\psi(x,D)|^2\approx \Big({Mv\over 2\pi \hbar D}\Big)
\Big|A\big(p=(xMv/D)\big)\Big|^2.
\end{equation}
For example, for diffraction through one slit having a width $w$, Eq.(11) 
reads 
\begin{equation}
A_{(1-slit)}(p)=
{1\over \sqrt{w}}\int_{-w/2}^{w/2}e^{-i(px^\prime )/\hbar }dx^\prime ,
\end{equation}
so that Eq.(13) yields the one slit diffraction pattern 
\begin{equation}
|\psi_{(1-slit)}(x)|^2\approx \Big({Mvw\sin^2(Mvwx/2\hbar D)
\over 2\pi \hbar D(Mvwx/2\hbar D)^2}\Big).
\end{equation}
For two slits, each of width $w$ and separated by a distance $b$ which 
obeys $b>w$, the momentum amplitudes obey 
\begin{equation}
A_{(2-slit)}(p)=\sqrt{2}\ \cos\Big({pb\over 2\hbar }\big)A_{(1-slit)}(p),
\end{equation}
leading to the diffraction pattern 
\begin{equation}
|\psi_{(2-slit)}(x)|^2\approx 2\cos^2\Big({Mvbx\over 2\hbar D}\big)
|\psi_{(1-slit)}(x)|^2.
\end{equation}
Thus, one sees that the momentum distribution near the slits 
give rise to the diffraction pattern in space near the detectors, 
via Eq.(13). Let us now turn to the quantum interference 
diffraction pattern expected from the momentum distribution 
of a trapped Boson condensate.

In the zero temperature Gross\cite{14}-Pitaevskii\cite{15} theory for dilute 
Bosons in a trap potential $U({\bf r})$, the variational trial ground state 
is taken to be ``quantum coherent'' in a sense that is closely analogous to 
coherent states in quantum optics. The trial ground state is a 
normalized eigenstate of the Boson field operator, 
\begin{equation}
\hat{\psi }({\bf r})|\Psi >=\Psi ({\bf r})|\Psi >.
\end{equation} 
The Bose condensation order parameter $\Psi ({\bf r})$ is then chosen to 
minimize the energy functional 
\begin{equation}
{\cal E}=<\Psi |\hat{H}-\mu \hat{N}|\Psi > 
\end{equation}
where $\mu $ is the chemical potential. Eq.(19) is evaluated using the 
two body scattering length $a$; 
\begin{equation}
{\cal E}=\int \Big\{\Psi^* (h-\mu)\Psi +
(2\pi \hbar^2 a/M)|\Psi |^4 \Big\}d^3r,
\end{equation}
where 
\begin{equation}
h=-\Big({\hbar ^2\over 2M}\Big)\nabla^2+U({\bf r}).
\end{equation}
Minimizing Eq.(20) yields  
\begin{equation}
h\Psi({\bf r})+(4\pi \hbar^2 a|\Psi ({\bf r})|^2/ M) 
\Psi ({\bf r}) =\mu \Psi ({\bf r}). 
\end{equation}

In the trapped Boson experiments, one starts the design of the trap 
with an inhomogeneous magnetic field so that $U_0({\bf r})$ 
behaves as an anisotropic oscillator potential with a single minimum 
at the origin. Such a single minimum in $U_0({\bf r})$ yields for the order 
parameter $\Psi_0 ({\bf r})$ a single maximum. This single maximum in 
$\Psi_0 ({\bf r})$ would behave as single source for the condensate explosion 
once the trap potential is removed. However, the single minimum in 
$U_0({\bf r})$ (i.e. the single maximum in $\Psi_0 ({\bf r})$) system is hard 
to build because the integer oriented spin atoms slip through a ``hole'' at 
the trapping potential minimum.

To fix this potential, some workers have plugged up the 
hole at the bottom of the magnetic trap with a LASER beam. The 
plug produces (as a side effect) an effective $U({\bf r})$ with 
{\em two minima}. Thus, traps with a plug 
can be (and have been) designed\cite{9} so that the order parameter 
$\Psi ({\bf r})$ has {\em two maxima}. In such a case, when the trap 
is removed and the condensate explodes away from the trap region, 
there will be a {\em double source} for the condensate Bosons. 
If these two sources  are {\em coherent}, then there was in the laboratory  
amplitude interference yielding order parameter fringes. These fringes
were known to be analogous to the interference fringes 
observed in conventional ``two slit diffraction''. If the two sources 
were {\em incoherent}, then such interference fringes would not have occured. 

It is to be expected that Bose condensed systems (even at 
finite temperature) will exhibit a well defined coherent order parameter 
$\Psi ({\bf r})$ whose condensate momentum distribution 
\begin{equation}
N_0({\bf p})=\Big|\int e^{-i({\bf p \cdot r})/\hbar}\Psi ({\bf r})
d^3r\Big|^2
\end{equation}
will properly reflect the two maxima in  $\Psi ({\bf r})$. Such a 
double maxima describing two coherent sources duplicate in 
$\rho ({\bf r},t)$ the interference pattern closely analogous 
to two slit diffraction. After the trapping potential is removed, the 
condensate atom contribution to the density is given by 
\begin{equation}
\rho_0 ({\bf r},t)\approx \int N_0({\bf p})
\delta \big({\bf r}-({\bf p}t/M)\big){d^3p\over (2\pi \hbar )^3}\ .
\end{equation} 
Reasoning as in Eqs.(16) and (17), and employing a separation vector 
${\bf b}$ which describes the position of the second peak in 
$\Psi ({\bf r})$ relative to the first peak in $\Psi ({\bf r})$, we 
find that 

\begin{equation}
N_{0(2-peak)}({\bf p})\approx I({\bf p},{\bf b})
N_{0(1-peak)}({\bf p}),
\end{equation}  
where the interference function 
\begin{equation}
I({\bf p},{\bf b})=\big(1+\eta \cos({\bf p\cdot b}/\hbar )\big),
\end{equation}
and $0<\eta<1$ describes the ``visibility'' of the interference fringes 
when the two peaks (i.e. sources) are only partially coherent.

Finally, we note that the oscillations in momentum ${\bf p}$ implicit 
in Eq.(26) for the interference function $I({\bf p},{\bf b})$, have appeared 
as oscillations in space ${\bf r}$ for a ``one frame snap shot'' \cite{9} 
of $\rho_0({\bf r},t_{fixed})$ fixed in time. Interference is also present 
as a time oscillation in $\rho_0({\bf r}_{fixed},t)$ for a movie picture of 
atoms in the neighborhood of a fixed point in space\cite{16}. 

The point is that the momentum, position and time are related by the 
``classical path rule'' ${\bf r}={\bf v}t=({\bf p}t/M)$, 
but with a {\em weighted momentum distribution}. This {\em essence} 
of the diffraction of particle beams as described by quantum mechanical 
amplitude superposition, should also be possible to observe in systems 
for which the one rotates the potential about an axis displaced from the 
symmetry axis. Mesoscopic quantum intereference should also be present in 
TOP traps\cite{7}.

\end{document}